\documentclass[12pt,letterpaper]{report}
\usepackage[margin=1in]{geometry}
\usepackage{pdfpages}
\usepackage{fancyhdr}
\usepackage{tocloft}
\usepackage{tikz}

\usepackage{hyperref}
\hypersetup{
    colorlinks,
    citecolor=black,
    filecolor=black,
    linkcolor=black,
    urlcolor=black
}

\renewcommand{\sectionmark}[1]{\markboth{#1}{}}
\newcommand{\fakesection}[1]{%
  \par\refstepcounter{section}
  \sectionmark{#1}
  \addcontentsline{toc}{section}{#1}
}

\renewcommand{\sectionmark}[1]{\markboth{#1}{}}

\renewcommand{\headrulewidth}{0pt}
\begin{document}

\title{Proceedings of the 2020 Scheme and Functional Programming Workshop}

\pagenumbering{roman}

\begin{titlepage}

  \font\myfont=cmr12 at 35pt
  \begin{minipage}{.6\textwidth}
    \hspace{-20pt}{\myfont Technical Report}
  \end{minipage}
  \begin{minipage}{.35\textwidth}
     \hspace{-5pt}\includegraphics[width=6cm]{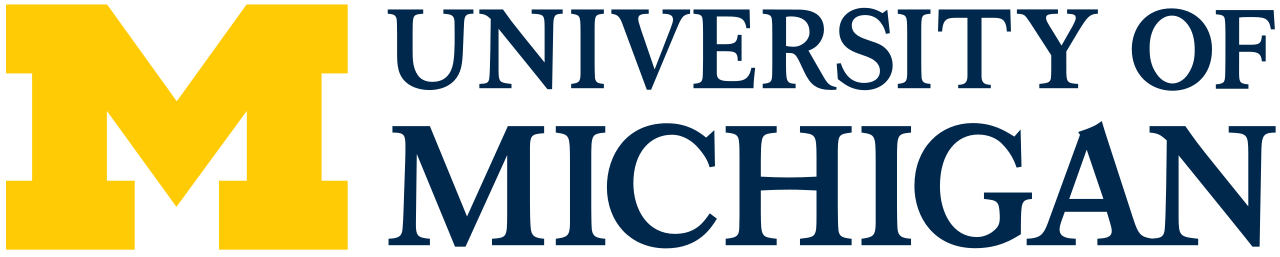}
  \end{minipage}
  \mbox{~}
  \noindent\makebox[\linewidth]{\rule{\textwidth}{3.0pt}}
  \vspace{5cm}
  \centering

	{\huge\bfseries Proceedings of the 2020 Scheme and Functional Programming Workshop\par}
  \vspace{3cm}
	{\LARGE Edited by Baptiste Saleil and Michael D. Adams\par}
  \vfill
	{\large January 2021\par}

\end{titlepage}

\clearpage
\newpage
\mbox{~}

\vfill
\noindent Computer Science and Engineering technical reports \\published by University of Michigan
are available at: \\

\def\UrlFont{\em}
\url{https://cse.engin.umich.edu/research/technical-reports-publications/}
\clearpage
\newpage
\mbox{~}

\pagestyle{fancy}
\renewcommand{\headrulewidth}{0pt}
\fancyhf{}
\rhead{}
\lhead{}
\rfoot{\thepage}
\lfoot{Scheme and Functional Programming Workshop 2020}

\section*{Preface}

This report aggregates the papers presented at the twenty-first annual Scheme and Functional Programming Workshop,
hosted on August 28th, 2020, online and co-located with the twenty-fifth International Conference on
Functional Programming.
The Scheme and Functional Programming Workshop is held every year to provide an opportunity for researchers and
practitioners using Scheme and related functional programming languages like Racket, Clojure, and Lisp, to share
research findings and discuss the future of the Scheme programming language.
Seven papers and three lightning talks were submitted to the workshop, and each submission was reviewed by three
members of the program committee. After deliberation, four papers and three lightning talks were accepted to the workshop.
In addition to the four papers and three lightning talks presented,
\begin{itemize}
  \item Martin Henz and Tobias Wrigstad gave an invited keynote speech entitled \textit{SICP JS: Ketchup on Caviar?}
  \item Bohdan Khomtchouk and Jonah Fleishhacker gave an invited keynote speech entitled \textit{21st Century Lisp in Academic Research and Pedagogy}.
\end{itemize}
Thank you to all the presenters, panelists, participants, and members of the program committee.

\vfill

\subsection*{Program Committee}

Michael D. Adams, University of Michigan (Program Co-Chair) \\
Baptiste Saleil, IBM Canada (Program Co-Chair) \\
Maxime Chevalier-Boisvert, Université de Montréal \\
Ryan Culpepper, Czech Technical University \\
Kimball Germane, University of Utah \\
Yukiyoshi Kameyama, University of Tsukuba \\
Andy Keep, Cisco Systems, Inc \\
Julien Pagès, Université de Montréal \\
Alexey Radul, Massachusetts Institute of Technology \\

\subsection*{Steering Committee}

Will Byrd, University of Alabama at Birmingham \\
Will Clinger, Northeastern University \\
Marc Feeley, Université de Montréal \\
Dan Friedman, Indiana University \\
Olin Shivers, Northeastern University \\

\clearpage
\newpage
\mbox{~}

\clearpage
\newpage

\tableofcontents
\thispagestyle{fancy}

\clearpage
\newpage

\pagestyle{fancy}
\renewcommand{\headrulewidth}{1pt}
\fancyhf{}
\rhead{}
\lhead{\rightmark}
\rfoot{\thepage}
\lfoot{Scheme and Functional Programming Workshop 2020}
\setlength{\headheight}{27.2pt}

\pagenumbering{arabic}

\fakesection{Keynote: SICP JS: Ketchup on Caviar?}
\lhead{Keynote - SICP JS: Ketchup on Caviar?}
\includepdf[pages=-,pagecommand={},width=\textwidth, trim=20mm 20mm 20mm 20mm]{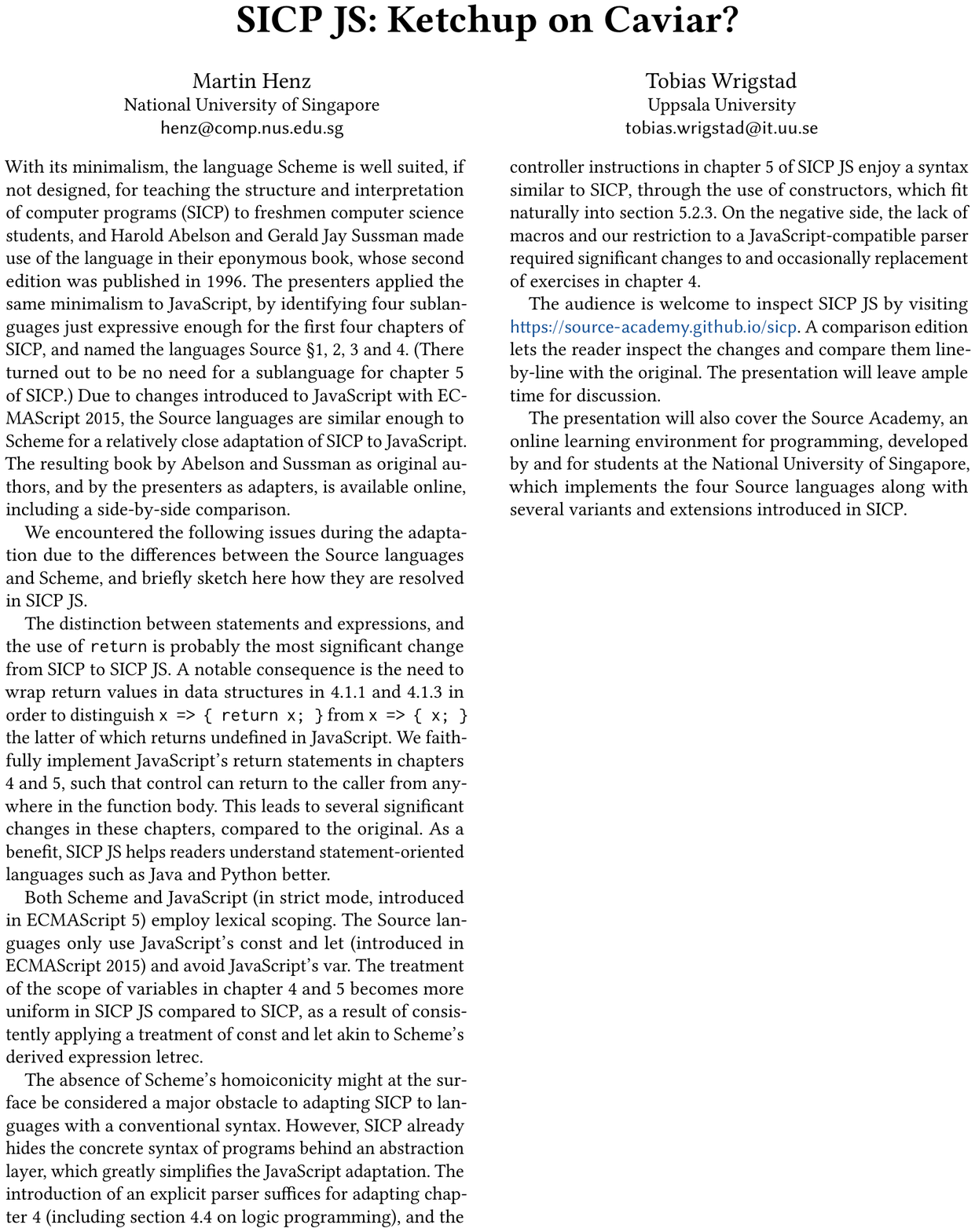}

\fakesection{Keynote: 21st Century Lisp in Academic Research and Pedagogy}
\lhead{Keynote - 21st Century Lisp in Academic Research and Pedagogy}
\includepdf[pages=-,pagecommand={}, width=\textwidth, trim=20mm 20mm 20mm 20mm]{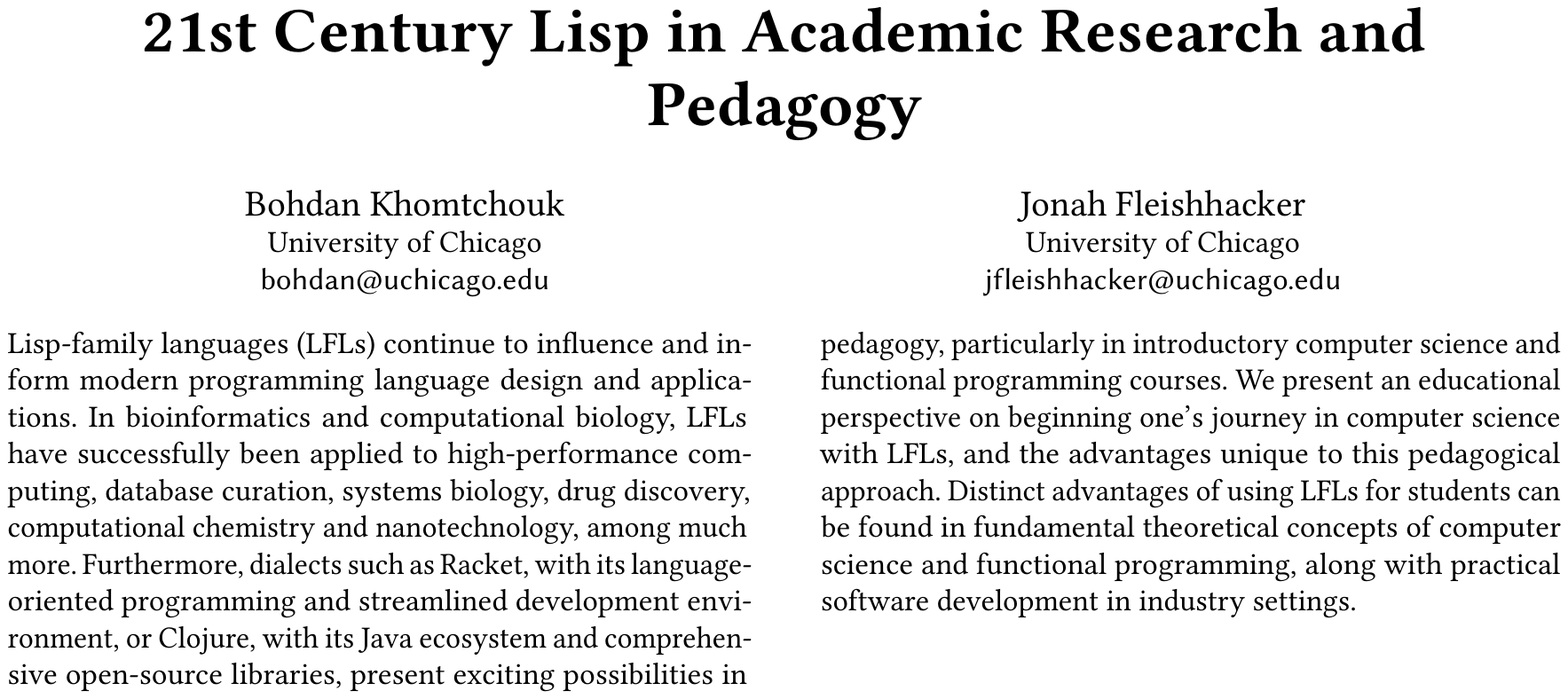}

\fakesection{Paper: Clotho: A Racket Library for Parametric Randomness}
\lhead{Paper - Clotho: A Racket Library for Parametric Randomness}
\includepdf[pages=-, width=\textwidth, trim=20mm 20mm 20mm 20mm,pagecommand={\begin{tikzpicture}[remember picture, overlay]
\fill[white!40!white] (-3,1.1) rectangle (18,0.7);
\end{tikzpicture}}]{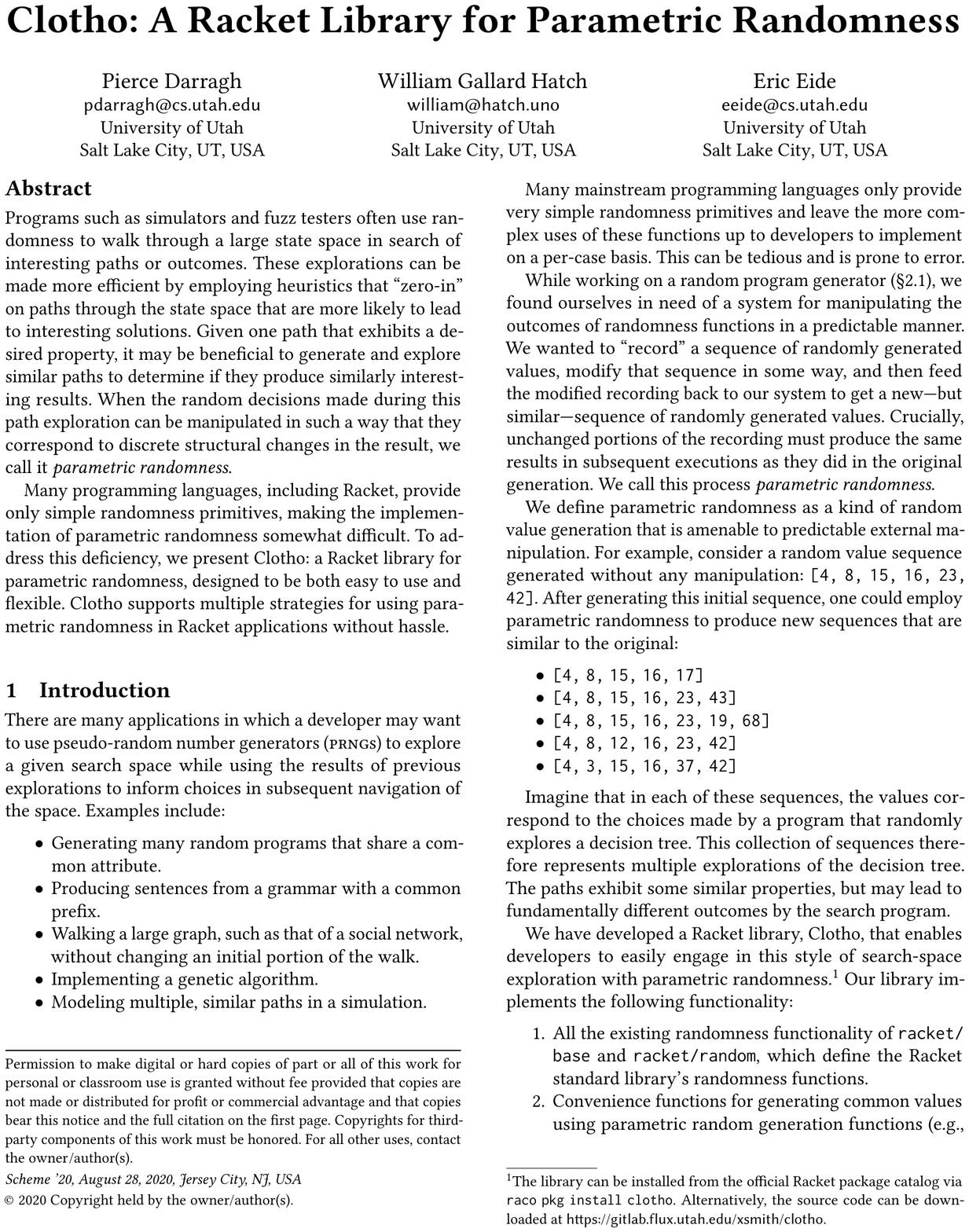}

\fakesection{Paper: Solving SICP: An Experience Report on Solving the World’s Most Famous Programming Problem Set}
\lhead{Paper - Solving SICP: An Experience Report on Solving the World’s Most Famous Programming Problem Set}
\includepdf[pages=-, width=\textwidth, trim=20mm 20mm 20mm 20mm,pagecommand={\begin{tikzpicture}[remember picture, overlay]
\fill[white!40!white] (-3,1.1) rectangle (18,0.7);
\fill[white!40!white] (-3,-22) rectangle (18,-21.2);
\end{tikzpicture}}]{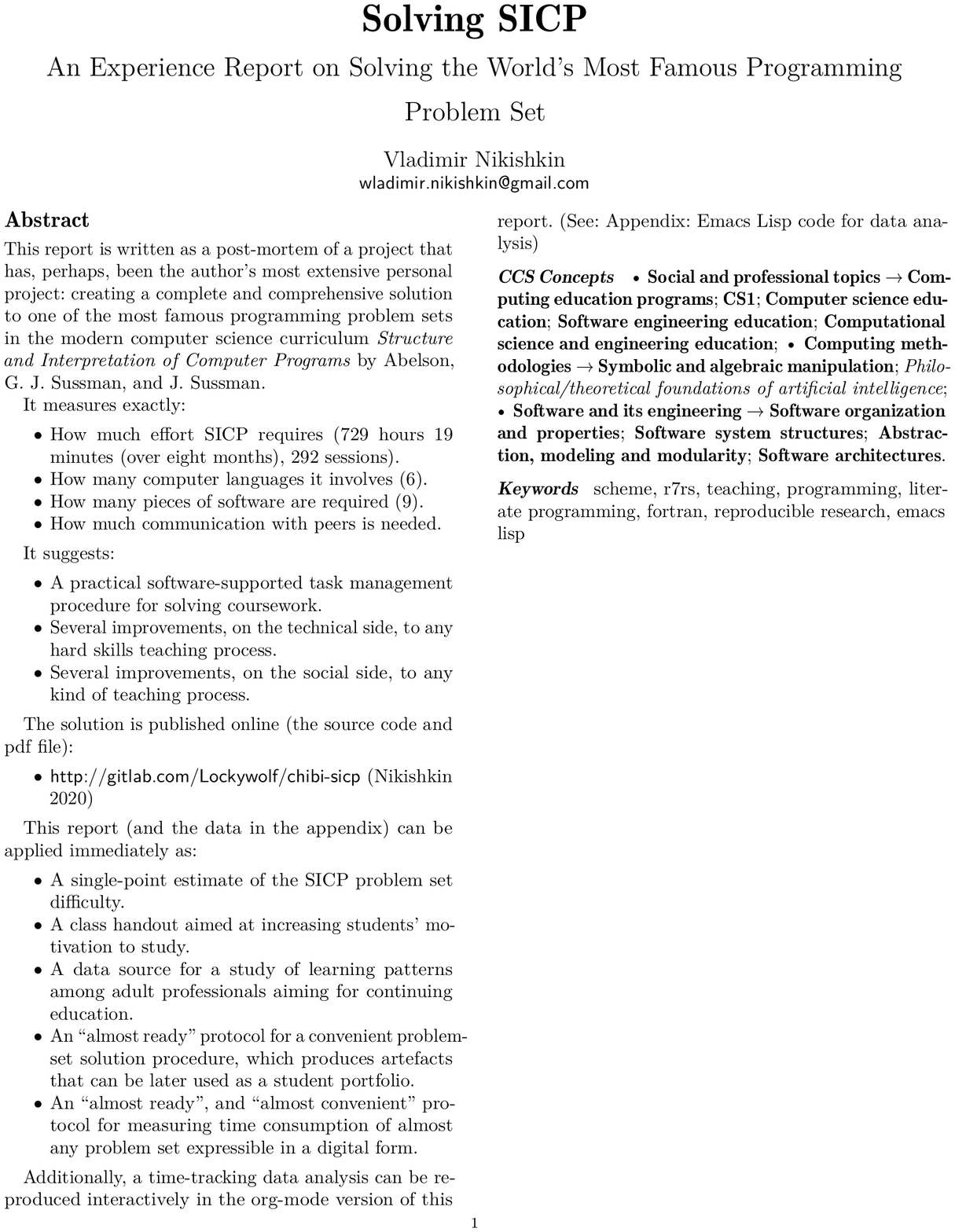}

\fakesection{Paper: Running Scheme On Bare Metal (Experience Report)}
\lhead{Paper - Running Scheme On Bare Metal (Experience Report)}
\includepdf[pages=-, width=\textwidth, trim=20mm 20mm 20mm 20mm,pagecommand={\begin{tikzpicture}[remember picture, overlay]
\fill[white!40!white] (-3,1.1) rectangle (18,0.7);
\fill[white!40!white] (-3,-22) rectangle (18,-21.2);
\end{tikzpicture}}]{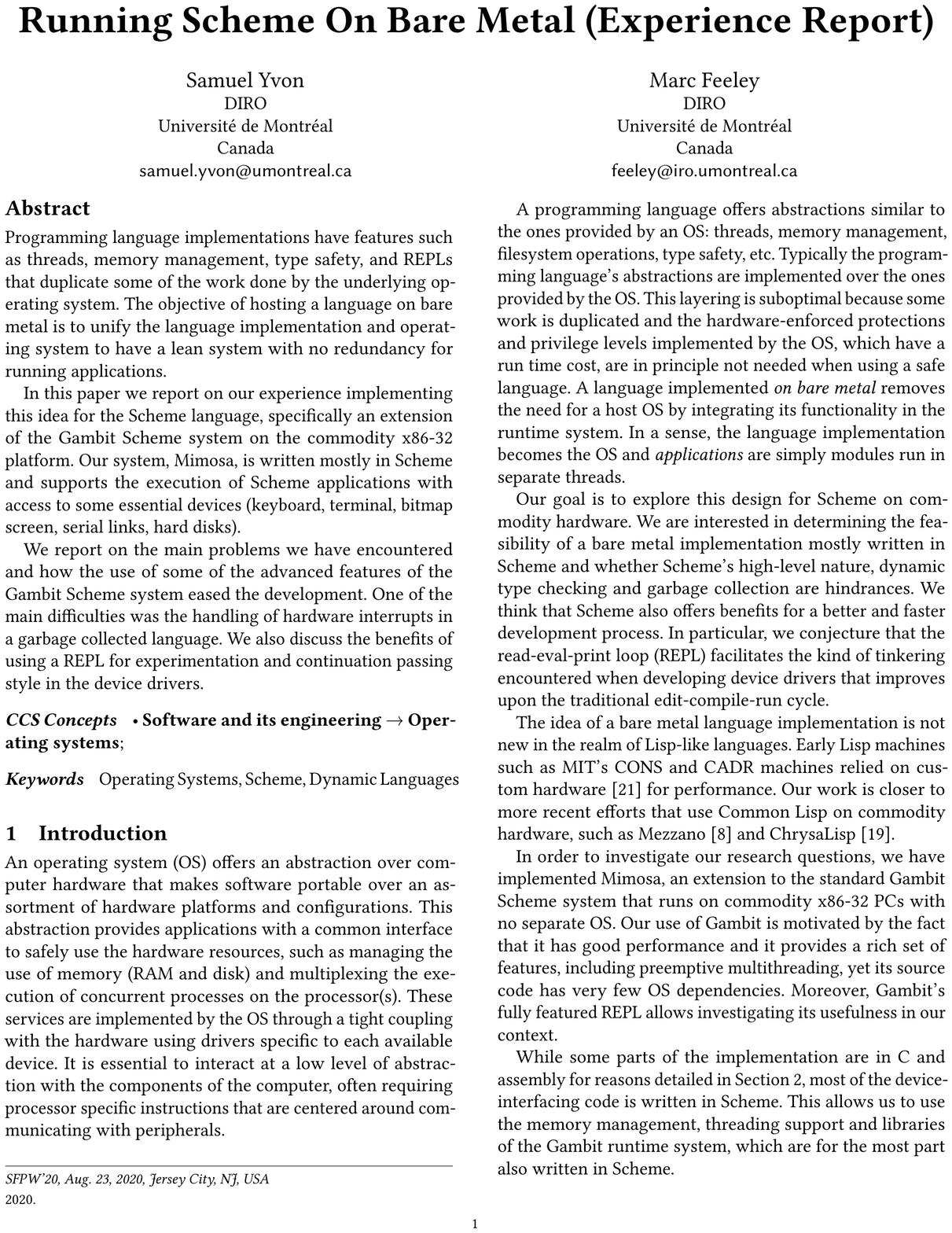}

\fakesection{Paper: Scheme for Scientific Computing}
\lhead{Paper - Scheme for Scientific Computing}
\includepdf[pages=-, width=\textwidth, trim=20mm 20mm 20mm 20mm,pagecommand={\begin{tikzpicture}[remember picture, overlay]
\fill[white!40!white] (-3,1.1) rectangle (18,0.7);
\end{tikzpicture}}]{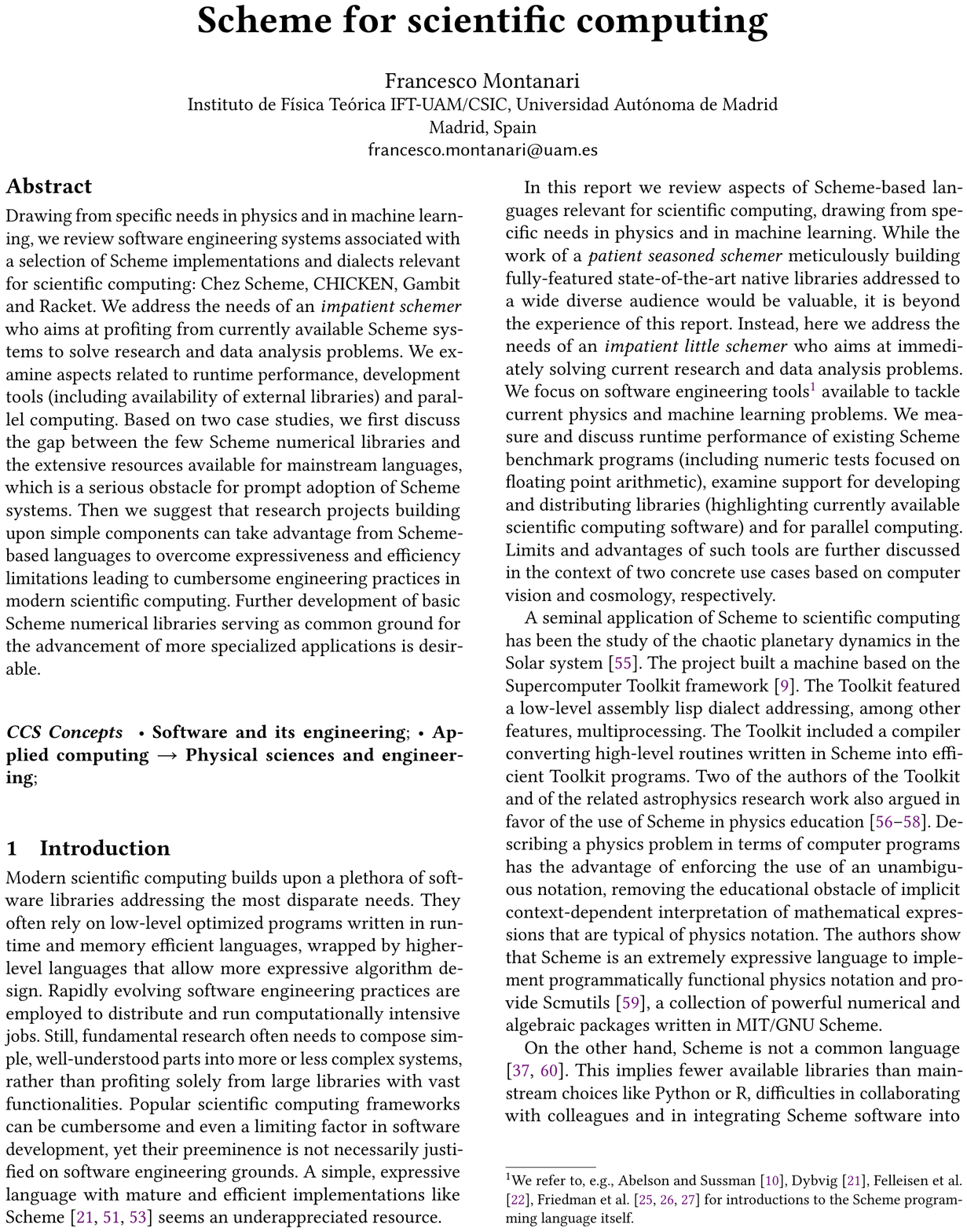}

\fakesection{Talk: On Teaching Type Systems as Macros}
\lhead{Talk - On Teaching Type Systems as Macros}
\includepdf[pages=-, width=\textwidth, trim=20mm 20mm 20mm 20mm,pagecommand={\begin{tikzpicture}[remember picture, overlay]
\end{tikzpicture}}]{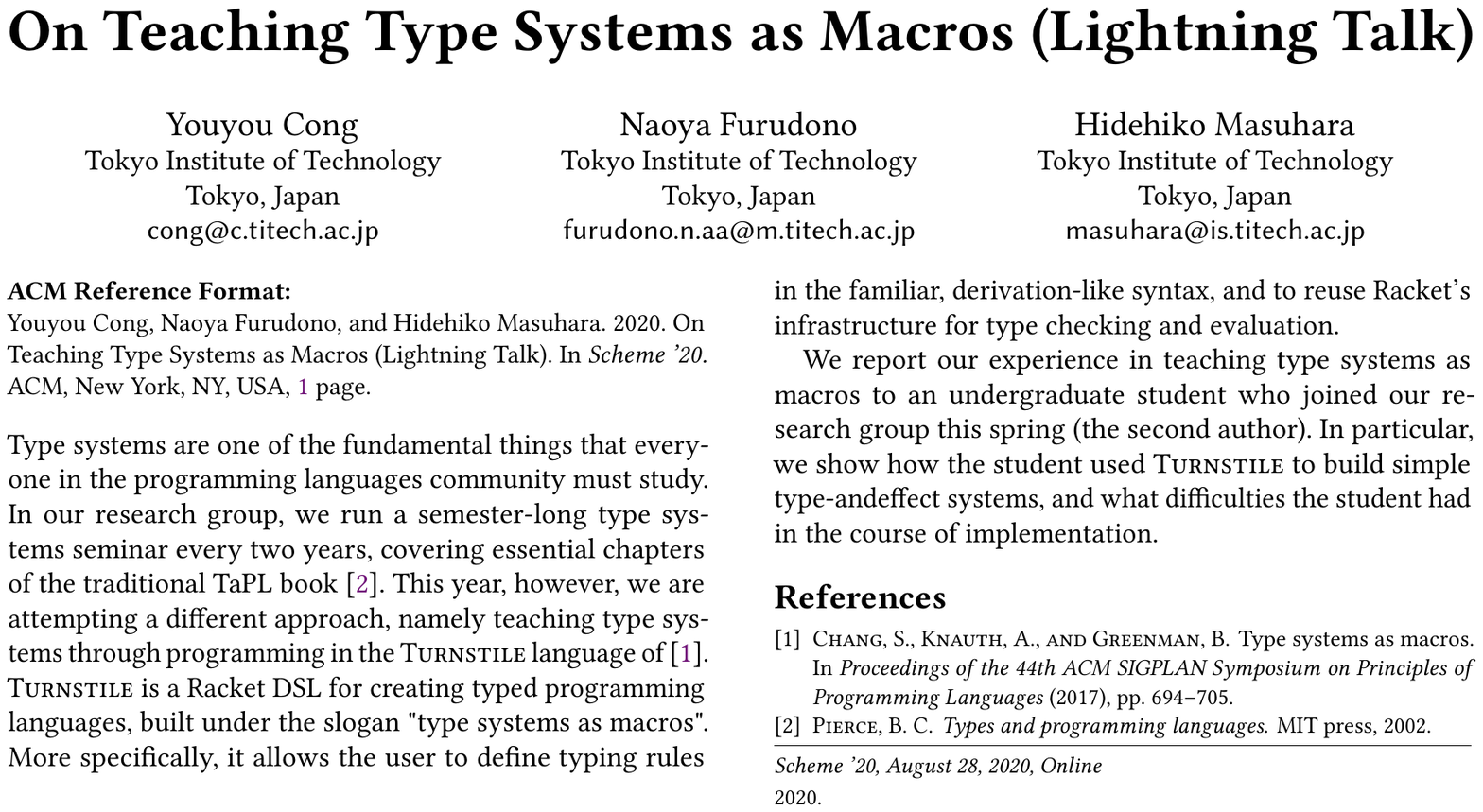}

\fakesection{Talk: Designing a Programming Environment Based on the Program Design Recipe}
\lhead{Talk - Designing a Programming Environment Based on the Program Design Recipe}
\includepdf[pages=-, width=\textwidth, trim=20mm 20mm 20mm 20mm,pagecommand={\begin{tikzpicture}[remember picture, overlay]
\fill[white!40!white] (7.5,-22) rectangle (7.8,-21);
\end{tikzpicture}}]{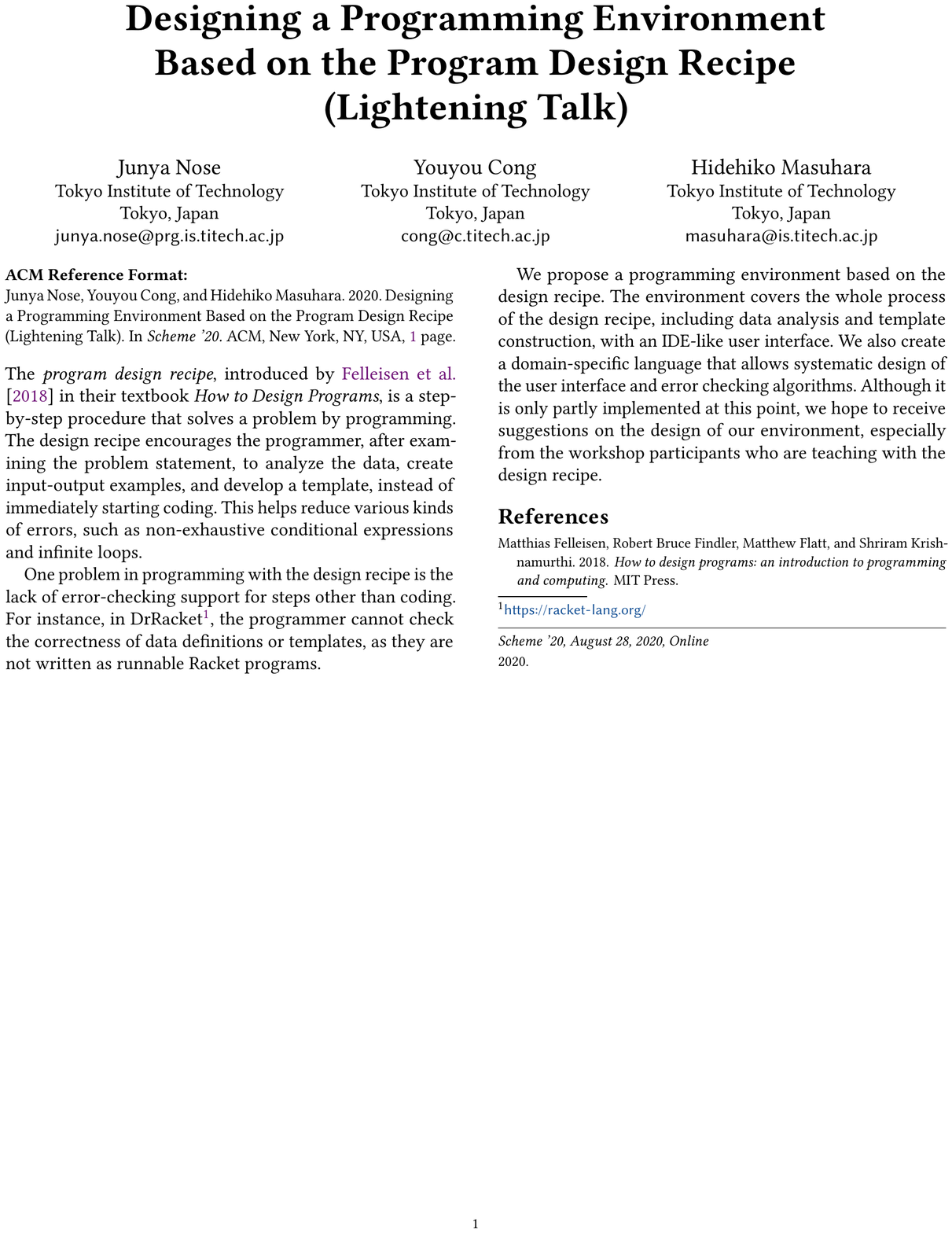}

\fakesection{Talk: Programming with Petri Nets to Reason about Concurrency}
\lhead{Talk - Programming with Petri Nets to Reason about Concurrency}
\includepdf[pages=-, width=\textwidth, trim=20mm 20mm 20mm 20mm,pagecommand={\begin{tikzpicture}[remember picture, overlay]
\fill[white!40!white] (-3,-19.8) rectangle (7.5,-22);
\end{tikzpicture}}]{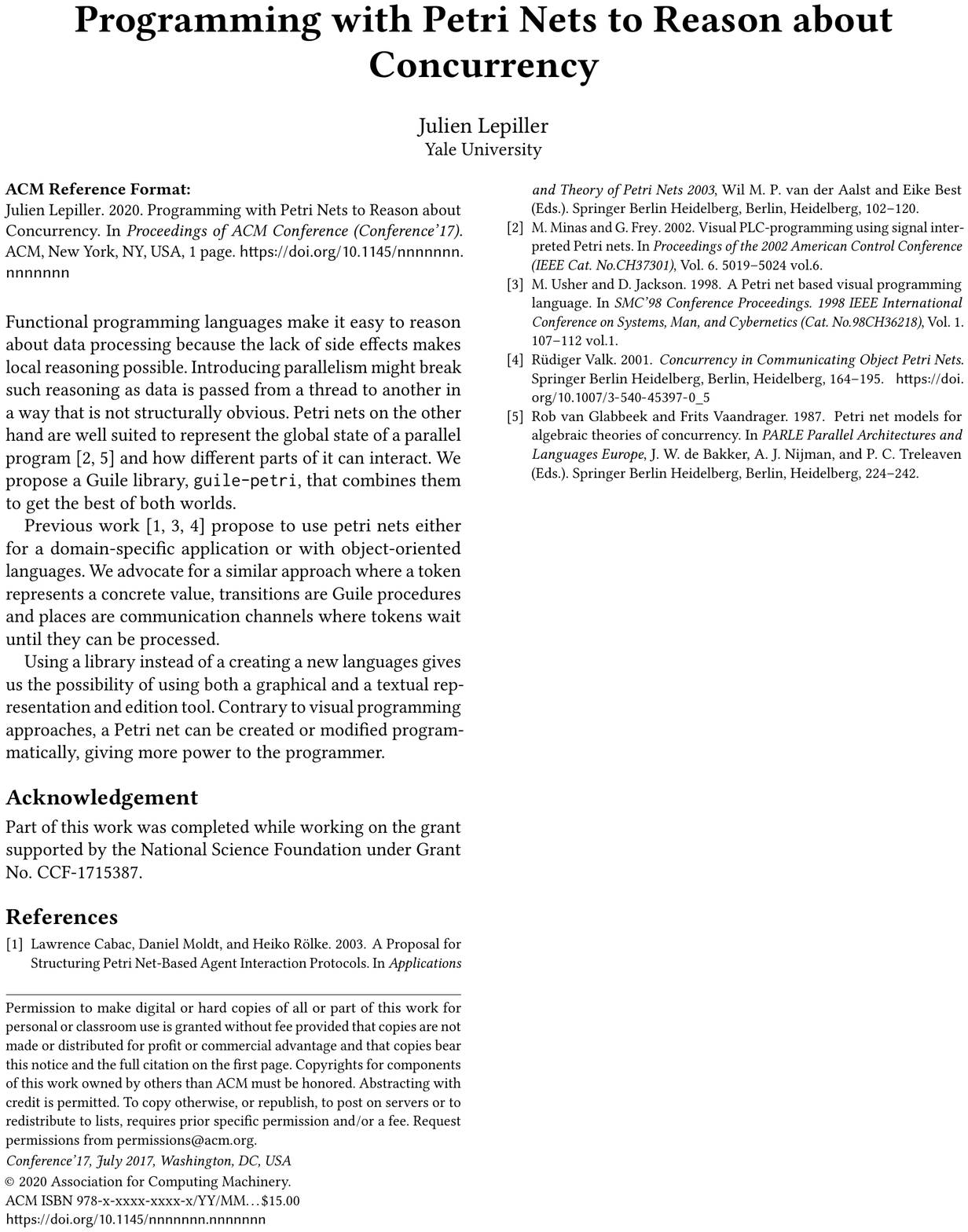}

\newpage

\newcommand{\authorentry}[2] { \begin{minipage}{.3\textwidth}#1\end{minipage}\begin{minipage}{.1\textwidth}\begin{flushright}#2\end{flushright}\end{minipage} }

\newcommand{\pl}[1] {\hyperlink{page.#1}{#1}}

\pagestyle{fancy}
\renewcommand{\headrulewidth}{0pt}
\fancyhf{}
\rhead{}
\lhead{}
\rfoot{\thepage}
\lfoot{Scheme and Functional Programming Workshop 2020}

\section*{Author Index}

\authorentry{Cong, Youyou}{\pl{77}, \pl{78}} \\
\authorentry{Darragh, Pierce}{\pl{3}} \\
\authorentry{Eide, Eric}{\pl{3}} \\
\authorentry{Feeley, Marc}{\pl{51}} \\
\authorentry{Fleishhacker, Jonah}{\pl{2}} \\
\authorentry{Furudono, Naoya}{\pl{77}} \\
\authorentry{Hatch, William G.}{\pl{3}} \\
\authorentry{Henz, Martin}{\pl{1}} \\
\authorentry{Khomtchouk, Bohdan}{\pl{2}} \\
\authorentry{Lepiller, Julien}{\pl{79}} \\
\authorentry{Masuhara, Hidehiko}{\pl{77}, \pl{78}} \\
\authorentry{Montanari, Francesco}{\pl{66}} \\
\authorentry{Nikishkin, Vladimir}{\pl{14}} \\
\authorentry{Nose, Junya}{\pl{78}} \\
\authorentry{Wrigstad, Tobias}{\pl{1}} \\
\authorentry{Yvon, Samuel}{\pl{51}} \\

\newpage

\end{document}